\documentclass{moriond}

\def\Journal#1#2#3#4{{#1} {\bf #2}, #3 (#4)}

\def\PRD{{\em Phys. Rev.} D}

\def\JHEP{{\em JHEP}}
\def\JINST{{\em JINST}}
\def\EPJC{{\em EPJC}}
\def\IJMP{{\em Int. J. Mod. Phys. A}}

\def\tev{\mbox{\,TeV}\xspace}
\def\invfb{\ensuremath{\mbox{\,fb}^{-1}}\xspace}
\def\pt{\mbox{$p_{\rm T}$}\xspace}
\def\ptsq{\mbox{$p^2_{\rm T}$}\xspace}

\def\be{\begin{equation}}
\def\ee{\end{equation}}
\def\bea{\begin{eqnarray}}
\def\eea{\end{eqnarray}}

\def\gevcc{\ensuremath{{\mathrm{\,Ge\kern -0.1em V\!/}c^2}}}
\def\gevc{\ensuremath{{\mathrm{\,Ge\kern -0.1em V\!/}c}}}
\usepackage{lineno,subfigure,xspace}
\newcommand{\Photo}{}

\begin{document}
\vspace*{4cm}
\title{ELECTROWEAK PHYSICS AND QCD AT LHCB}

\author{ D. JOHNSON\\
On behalf of the LHCb collaboration }

\address{CERN, Geneva, Switzerland}

\maketitle\abstracts{
Data collected by the LHCb experiment allow proton structure functions to be probed in a kinematic region beyond the reach of other experiments, both at the LHC and further afield. In these proceedings the significant impact of LHCb Run 1 measurements on PDF fits is recalled and recent LHCb results, that are sensitive to PDFs, are described.}

\section{Measuring parton distribution functions at LHCb}
The current experimental knowledge of proton structure functions carries large uncertainties, which are often the limiting factor in theoretical predictions for Standard Model (SM) processes and those beyond it (BSM). This strongly motivates the continued experimental study of processes that provide sensitivity to parton distribution functions (PDFs).

PDFs are typically parameterised using two kinematic variables, $Q^2$ and Bjorken $x$, corresponding to the squared interaction energy and the fraction of the nucleon's momentum carried by the parton, respectively. By consequence of the forward LHCb detector geometry, results obtained using LHCb data allow constraints to be placed on PDFs in kinematic regions that are inaccessible to those from other experiments at HERA and at the LHC. 

A significant effect is seen in PDF fits~\cite{PDFconstraints} even when employing only a small selection of LHCb Run 1 results. These proceedings describe three recently published studies of heavy vector boson production at LHCb which will constrain PDF fits further. In addition, a new analysis of $\Upsilon$ photoproduction is presented, with particular sensitivity to the gluon PDF at low $x$.

\section{Forward production of heavy vector bosons in 7 and 8\tev $pp$ collisions}
Three analyses of $pp$ collisions in LHCb are considered: $W$ boson production at 7\tev~\cite{LHCbWprod7TeV}; $Z$ boson production, reconstructed in the $e^+e^-$ decay mode at 8\tev~\cite{LHCbZ2eeprod8TeV}; and 7\tev $Z$ boson production associated with a $b$-jet, where the $Z$ decays to the dimuon final state~\cite{LHCbZplusbprod7TeV}. There are several common components to these analyses. Final state muons and electrons are required to lie in the pseudorapidity range between 2 and 4.5 and to have a \pt greater than 20\gevcc. In addition, most of these measurements benefit from a very precise integrated luminosity measurement~\cite{LHCbIntLumi}, with a relative precision of 1.7\% for 7\tev data and 1.2\% for 8\tev data~\footnote{The $Z+b$-jet production cross-section relies on an earlier determination, with 3.5\% relative uncertainty.}.

The first study exploits the full 7\tev $pp$ LHCb data sample corresponding to an integrated luminosity of 1\invfb. A single, high-\pt muon is required to be reconstructed. It must be well isolated, reducing background from hadronic QCD processes, and have a small impact parameter with respect to the primary vertex, suppressing background from leptonic $\tau$ and semi-leptonic heavy flavour decays. Events containing a second muon with large transverse momentum (\pt) are rejected in order to reduce $\gamma^*/Z\to\mu^+\mu^-$ background. The final sample contains 806,094 $W^\pm$ candidates. A fit to the candidate \pt distribution is carried out in eight pseudorapidity ranges in order to identify remaining contributions from muons originating in $K^\pm$ or $\pi^\pm$ decays, other electroweak boson decays or from decaying heavy flavour, and the overall sample purity is found to be approximately 77\%. The tag-and-probe method is employed to determine reconstruction efficiencies from samples of $Z\to\mu^+\mu^-$ decays, which are also used to determine selection efficiencies, after correcting for the difference in muon \pt. The integrated luminosity uncertainty dominates the systematic uncertainty for the $\sigma_{W\to\mu\nu}$ cross-section measurements, and is followed by the uncertainty on the muon reconstruction efficiencies. Both of these uncertainties cancel in the ratio, however, and in that case the main source of uncertainty arises from the template descriptions used in the fit to candidate \pt.

\begin{figure}[htbp]
\centering
\subfigure[Total cross-section.]{\includegraphics[width=.32\textwidth]{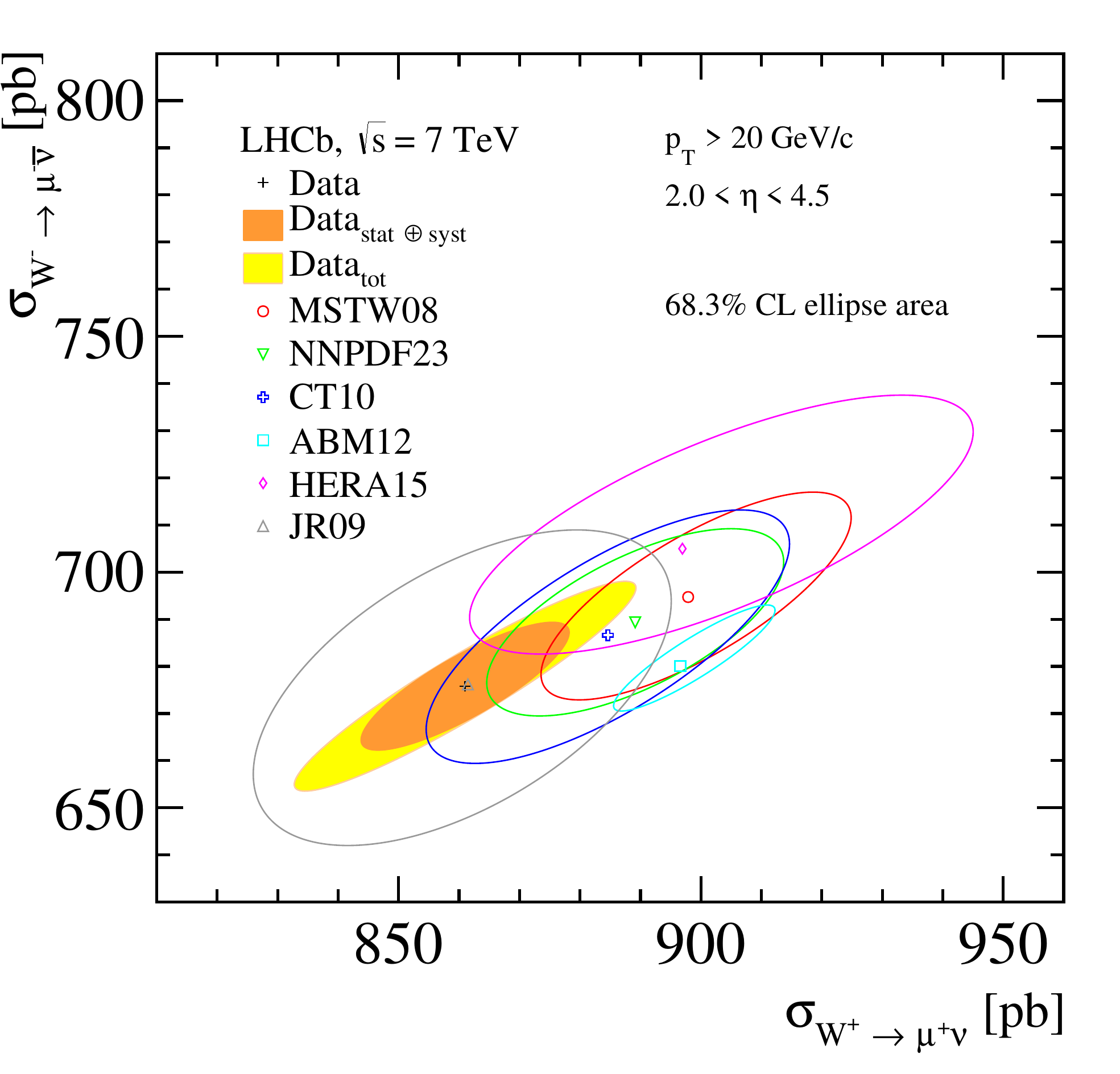}\label{im:TotW}}
\subfigure[Differential cross-section.]{\includegraphics[width=.45\textwidth]{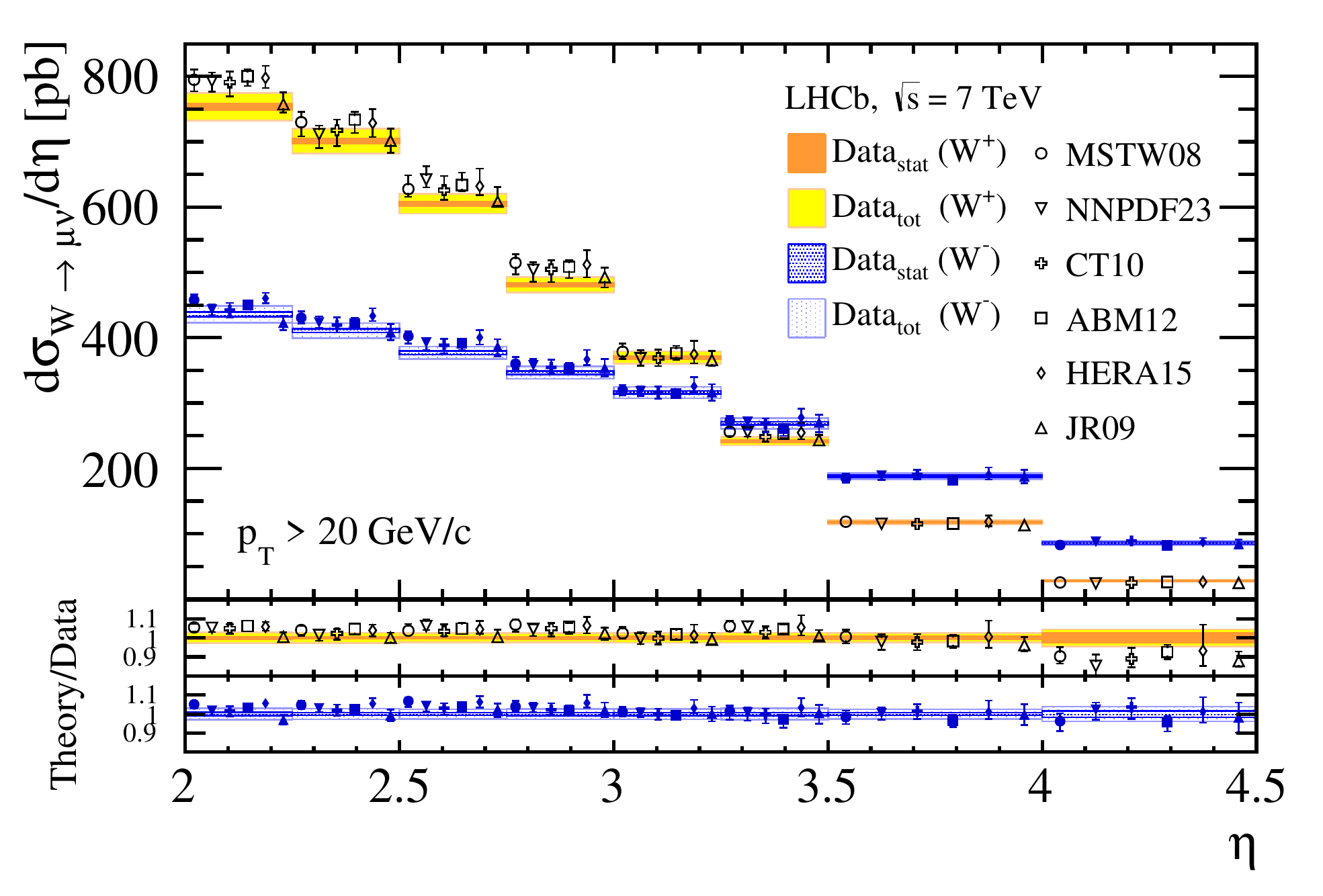}\label{im:DiffW}}
\caption{Total (a) and differential (b) cross-section measurements for $W^+$ and $W^-$ production in $pp$ collisions at 7\tev.}
\end{figure}

The total cross-sections (Figure~\ref{im:TotW}) and differential cross-sections (Figure~\ref{im:DiffW}) demonstrate good agreement between the data and a range of SM next-to-next-to-leading order (NNLO) calculations.

The second result described concerns $Z$ boson production reconstructed in the $Z\to e^+e^-$ channel, using the 8\tev $pp$ data set which corresponds to an integrated luminosity of 2\invfb. This result extends a previous previous analysis using only 7\tev $pp$ collisions~\cite{7tevZ2ee} and is an attractive extension to the $Z\to\mu^+\mu^-$ channel, offering a statistically independent measurement with considerably different systematic uncertainties. The 65,552 $Z\to e^+e^-$ candidates are selected requiring that the electrons should have a well-reconstructed momentum in the spectrometer and be associated to tracks leaving significant deposits in the electromagnetic calorimeter but not in the hadronic calorimeter. Nevertheless the principal background arises from randomly combined misidentified hadrons and is modelled using a sample of same-sign electron pairs reconstructed in data. 

Since the electrons pass through a significant amount of detector material, and the effect of irrecoverable bremsstrahlung occurring before the electrons pass the magnet is considerable, there is significant uncertainty on the measurement of their momenta. Although some energy correction is achieved using associated calorimeter deposits, momenta are still degraded by approximately 25\%. Consequently, the result is presented in terms an angular variable $\phi^*\approx\pt/M$ and boson rapidity, $y_Z$, instead of \pt and $y_Z$. The largest systematic uncertainty, 1.6\%, arises from the limited knowledge of electron tracking efficiencies and is obtained by comparing the efficiencies observed in simulation and data. 

\begin{figure}[htbp]
\centering
\subfigure[Differential cross-section as a function of $y_Z$.]{\includegraphics[width=.32\textwidth]{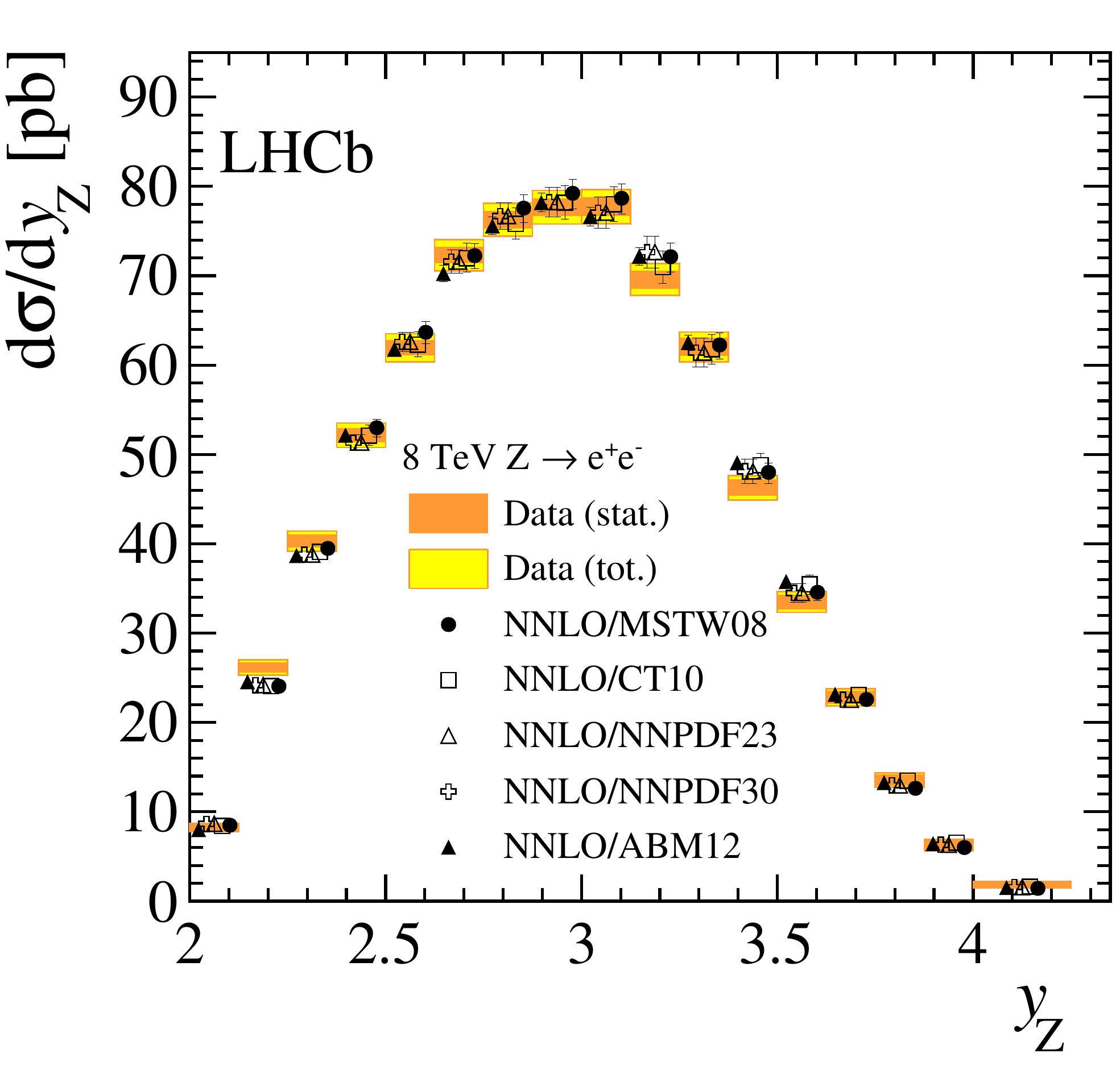}\label{im:Z2ee_yZ}}
\hspace{.5cm}
\subfigure[Ratio of predicted and measured cross-section as a function of $\phi^*$.]{\includegraphics[width=.32\textwidth]{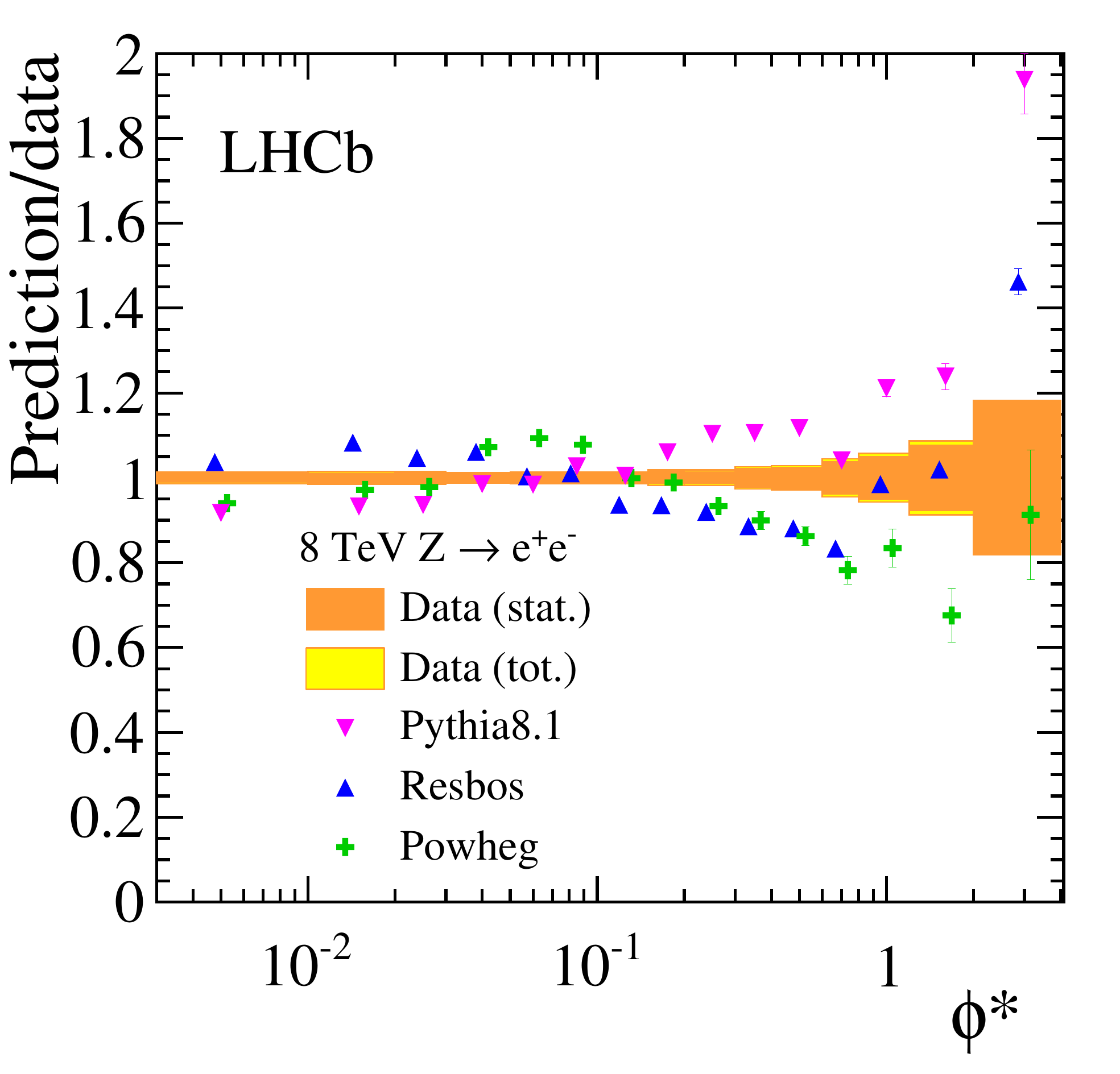}\label{im:Z2ee_phistar}}
\caption{Differential $Z$ boson production cross-sections in 8\tev $pp$ collisions, reconstructed in the $Z\to e^+e^-$ channel. In (a) the differential production cross-section is shown as a function of $Z$ rapidity whereas in (b) the ratio of prediction to measurement is shown as a function of the angular variable $\phi^*\approx\pt/M$.\label{im:Z2ee}}
\end{figure}

The measured differential cross-sections are shown in Figure~\ref{im:Z2ee}. Figure~\ref{im:Z2ee_yZ} demonstrates the good agreement of theoretical predictions with LHCb data, down to Bjorken-$x\approx10^{-6}$, for a range of PDF sets. In Figure~\ref{im:Z2ee_phistar} the agreement is evident between theoretical predictions using different methods to treat soft gluon emissions~\cite{Pythia,Powheg,Resbos}.

\begin{figure}[htb]
\centering
\subfigure[Fit to SV corrected mass.]{\includegraphics[width=.32\textwidth]{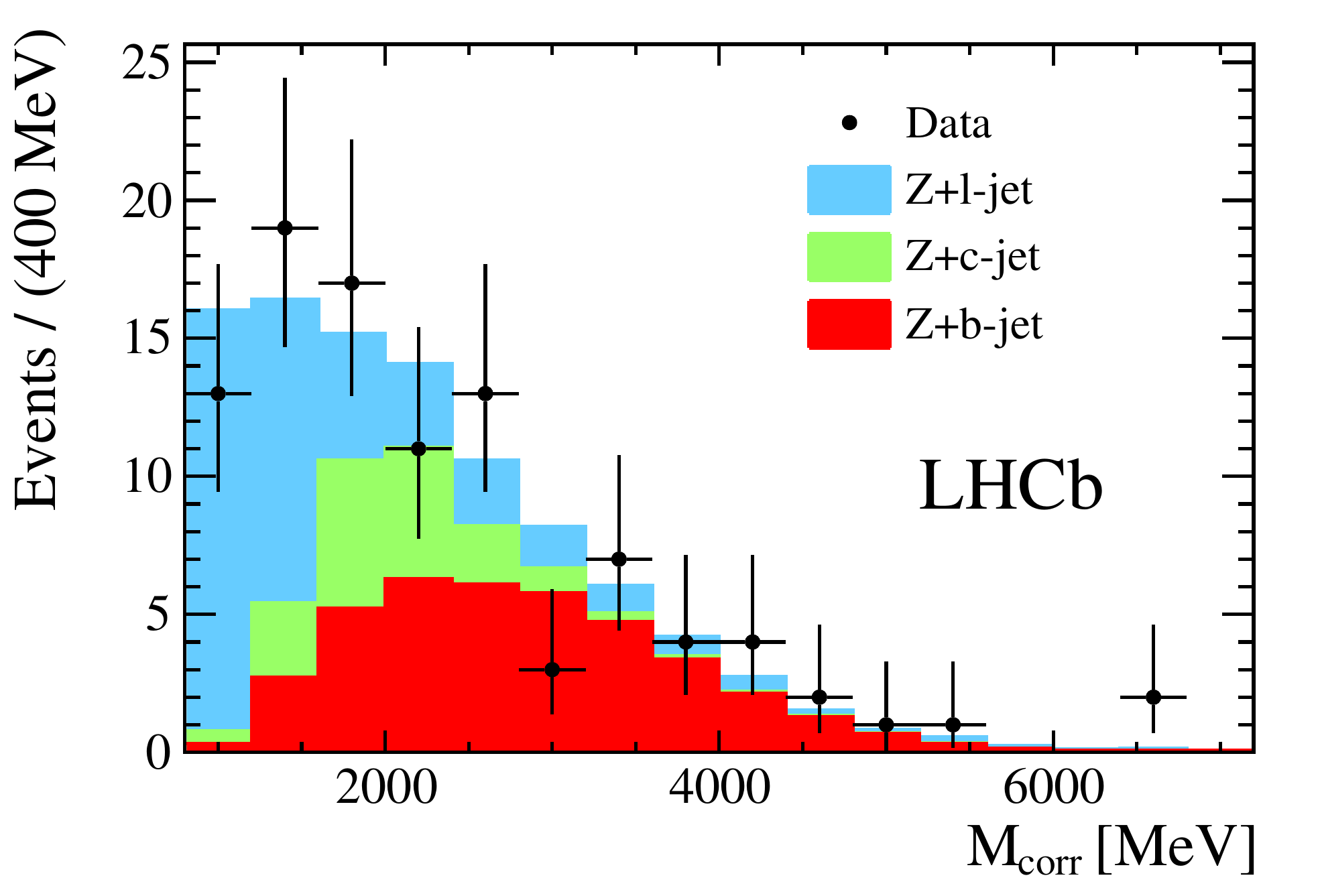}\label{im:Z+bMcorr}}
\hspace{.5cm}
\subfigure[Cross-section.]{\includegraphics[width=.365\textwidth]{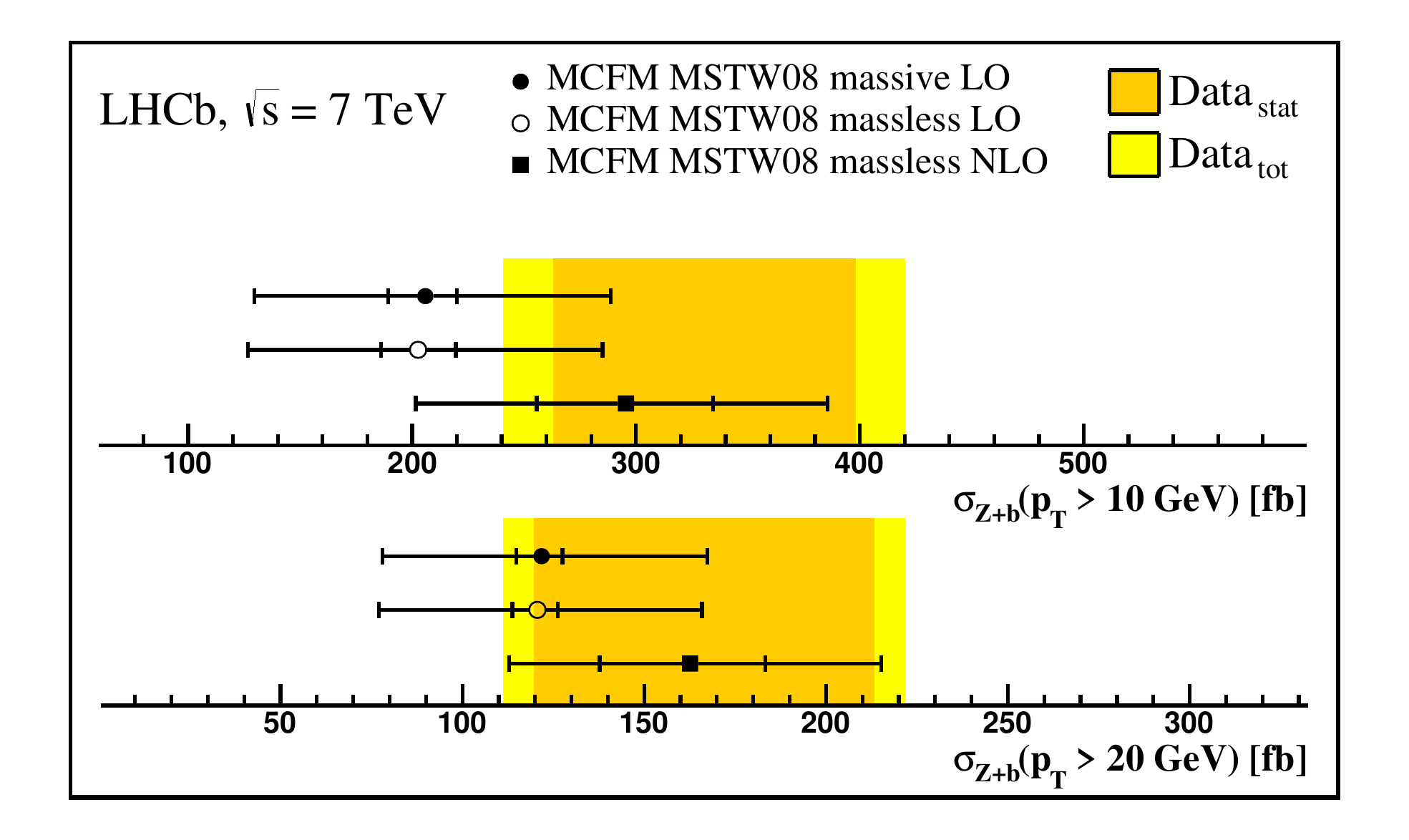}\label{im:Z+bCrossSection}}
\caption{Separation between $Z$ production in association with light, charm and beauty jets is obtained through a fit to the invariant mass of a secondary vertex reconstructed within the jet, shown in (a). The resulting $Z+b$-jet production cross-section is compared to various theoretical predictions in (b) and good agreement is seen.}
\end{figure}

The third study concerns the production of $Z$ bosons in association with a beauty quark and uses data collected in 7\tev $pp$ collisions, corresponding to an integrated luminosity of 1\invfb. The measurement is normalised to the earlier LHCb measurement of $Z$ boson production in association with a light jet~\cite{ZplusLightJet} and forms a benchmark measurement to constrain backgrounds in SM Higgs analyses and BSM searches. The $Z$ boson is reconstructed through its decay to $\mu^+\mu^-$ in the same way as the $Z$ in the first analysis described in these proceedings. The associated jet is reconstructed using a particle flow algorithm with the following inputs: charged tracks and calorimeter clusters for neutral particles where deposits associated with charged tracks are subtracted. An anti-$k_{\rm T}$ clustering algorithm is employed with distance parameter equal to 0.5. The identification of $b$-jets is achieved by searching for 2-, 3- and 4-particle secondary vertices (SV) within the reconstructed jet, with the characteristic topology and kinematics expected for $b$-hadron decays. A fit, shown in Figure~\ref{im:Z+bMcorr}, is performed to the corrected invariant-mass of the secondary vertex using templates for light, charm and beauty jets, and $72\pm15$ $Z+b$-jet candidates are identified with jet \pt$>10\gevc$. The reconstruction efficiencies obtained in the $Z$+jet analysis are applied to this study, and the extra $b$-tagging efficiency is studied using simulated samples. This additional efficiency determination is the largest source of systematic uncertainty, along with that arising from the fit to SV corrected invariant-mass.

The measured cross-section is shown in Figure~\ref{im:Z+bCrossSection}. Comparison is made with MCFM calculations~\cite{MCFM} using massless and massive $b$-quarks at various orders of perturbative expansion, and good agreement with the data is seen in each case.

\section{Exclusive production of heavy flavour states in diffractive interactions}
Exclusive photoproduction occurs when a state is produced by low-\pt, colourless exchange of a pomeron and photon between the incoming protons, without dissociation of the protons or additional gluon radiation. Such processes can be perturbatively calculable and are sensitive to the square of the gluon PDF down to Bjorken-$x$ of $1.5\times10^{-5}$. Since the photon-proton interaction energy depends exponentially on the produced meson rapidity, the forward geometry of LHCb allows the exploration of a new kinematic region compared to earlier measurements of exclusive $\Upsilon$ photoproduction at HERA.

\begin{figure}[htb]
\centering
\subfigure[Differential cross-section]{\includegraphics[width=.32\textwidth]{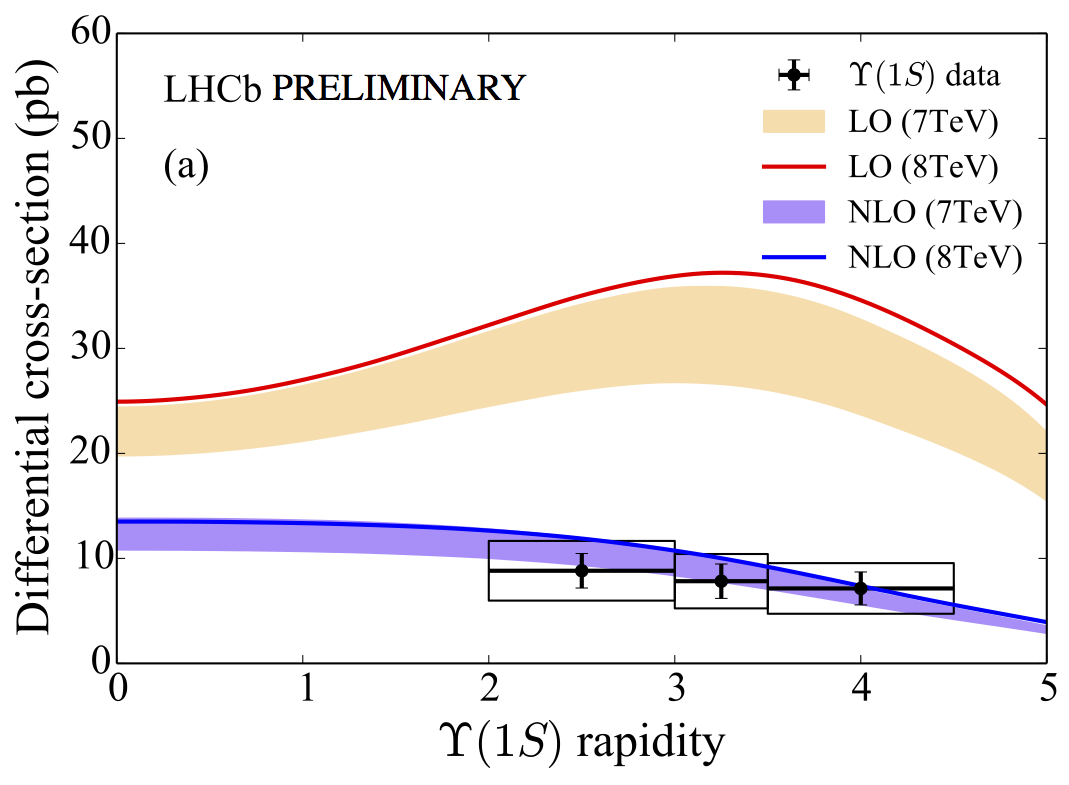}\label{im:UpsilonCrossSection}}
\hspace{.5cm}
\subfigure[Photo-production cross-section]{\includegraphics[width=.33\textwidth]{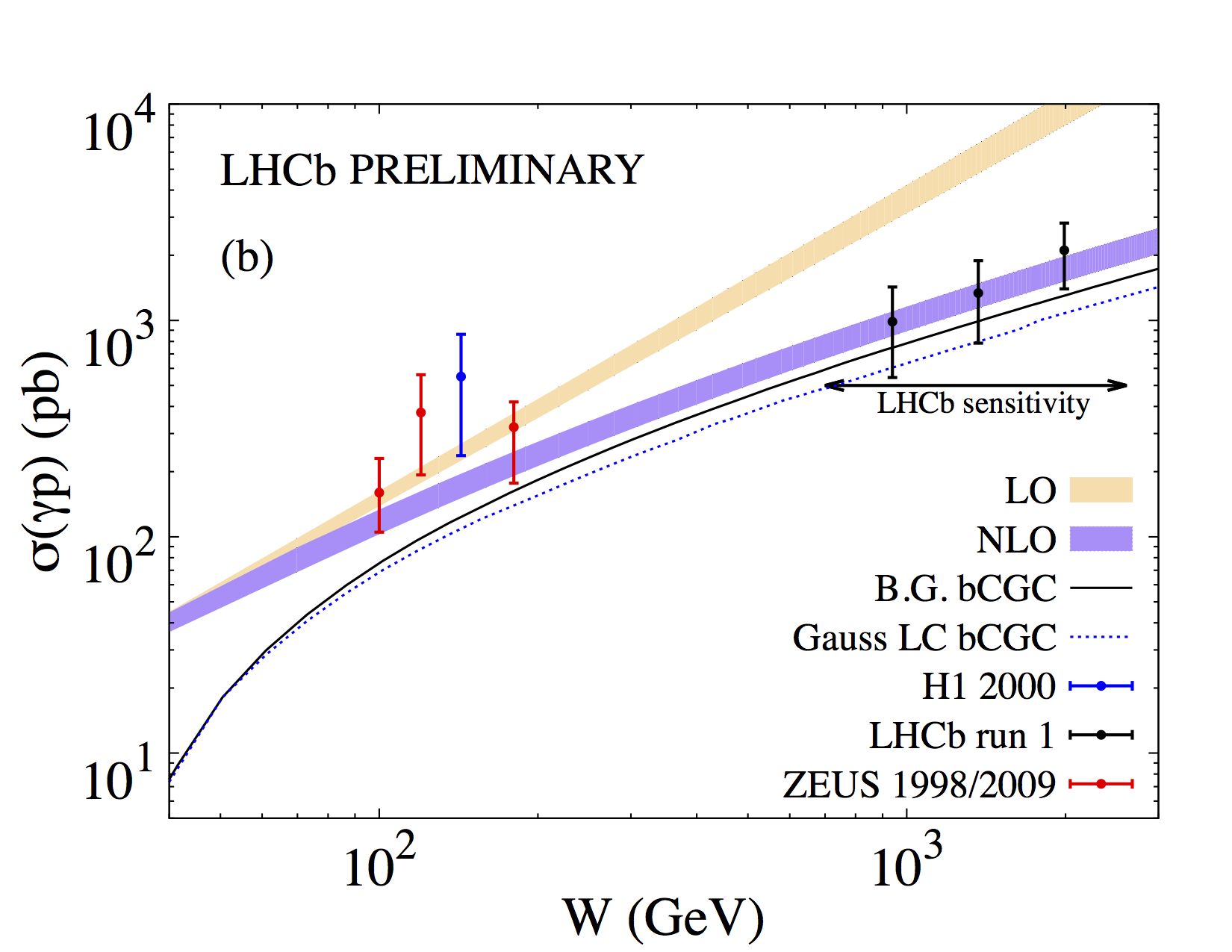}\label{im:UpsilonCrossSection_W}}
\caption{The differential cross-section as a function of rapidity, measured for $\Upsilon(1S)$ photoproduction, is shown in (a) and the derived photoproduction cross-section as a function of photon-proton interaction energy, $W$, is shown.}
\end{figure}

The experimental signature of these processes is simple: two well-reconstructed muons and little other detector activity are required. Background from radiative $\chi_b$ decays are quantified in data. The background from inclusive $\Upsilon$ production where the additional activity occurs outside the LHCb acceptance is quantified by fitting the $\Upsilon$ $\ptsq$ spectrum, for which exclusive production templates are derived using the SuperChiC generator~\cite{SuperChiC}. This fit is the source of the largest systematic uncertainties. These will be reduced in Run 2 by virtue of the recent addition of forward shower counters, extending the pseudorapidity coverage around LHCb~\cite{Herschel}. The measured production cross-section for the $\Upsilon(1S)$ is shown in Figure~\ref{im:UpsilonCrossSection} and the derived photo-production cross-section is given in Figure~\ref{im:UpsilonCrossSection_W}. Good agreement with perturbative predictions made at next-to-leading order is observed and the new kinematic region probed with the LHCb result is indicated.

\end{document}